\documentclass[prd,nofootinbib,showpacs,preprint]{revtex4}
\usepackage{amsmath}
\usepackage{graphicx}
\graphicspath{{Figs/}}
\usepackage{dcolumn}
\usepackage{bm}
\usepackage{amssymb}
\usepackage[usenames,dvipsnames]{color}
\usepackage{slashed}
\usepackage[dvipdfm,colorlinks,citecolor=blue]{hyperref}
\begin{document}
\title{Neutrino Masses and Leptogenesis in Type I+II Seesaw Models}
\author{Debasish Borah}
\email{dborah@tezu.ernet.in}
\affiliation{Department of Physics, Tezpur University, Tezpur-784028, India}
\author{Mrinal Kumar Das}
\email{mkdas@tezu.ernet.in}
\affiliation{Department of Physics, Tezpur University, Tezpur-784028, India}


\begin{abstract}
The baryon to photon ratio in the present Universe is very accurately measured to be $(6.065 \pm 0.090) \times 10^{-10}$. We study the possible origin of this baryon asymmetry in the neutrino sector through the generic mechanism of baryogenesis through leptogenesis. We consider both type I and type II seesaw origin of neutrino masses within the framework of left right symmetric models (LRSM). Using the latest best fit global neutrino oscillation data of mass squared differences, mixing angles and Dirac CP phase, we compute the predictions for baryon to photon ratio keeping the Majorana CP phases as free parameters for two different choices of lightest neutrino mass eigenvalue for both normal and inverted hierarchical patterns of neutrino masses. We do our calculation with and without lepton flavor effects being taken into account. We choose different diagonal Dirac neutrino mass matrix for different flavor effects in such a way that the lightest right handed neutrino mass is in the appropriate range. We also study the predictions for baryon asymmetry when the neutrino masses arise from a combination of both type I and type II seesaw (with dominating type I term) and discriminate between several combinations of Dirac and Majorana CP phases by demanding successful predictions for baryon asymmetry.
\end{abstract}
\pacs{12.60.-i,12.60.Cn,14.60.Pq}
\maketitle
\section{Introduction}
The fact that neutrinos have non-zero but tiny masses \cite{PDG} compared to other fermions in the Standard Model have been verified adequately in the recent neutrino oscillation experiments. The relative smallness of three Standard Model
neutrino masses can be naturally explained 
via seesaw mechanism which can be of three types: type I \cite{ti}, type II \cite{tii} and type III \cite{tiii}. All these mechanisms involve the inclusion of additional fermionic or scalar fields to generate tiny neutrino masses at tree level. Recent neutrino oscillation experiments T2K \cite{T2K}, Double ChooZ \cite{chooz}, Daya-Bay \cite{daya} and RENO \cite{reno} have not only made the earlier predictions for neutrino parameters more precise, but also measured a non-zero value of the reactor mixing angle $\theta_{13}$. The latest global fit value for $3\sigma$ range of neutrino oscillation parameters \cite{schwetz12} are as follows:
$$ \Delta m_{21}^2=(7.00-8.09) \times 10^{-5} \; \text{eV}^2$$
$$ \Delta m_{31}^2 \;(\text{NH}) =(2.27-2.69)\times 10^{-3} \; \text{eV}^2 $$
$$ \Delta m_{23}^2 \;(\text{IH}) =(2.24-2.65)\times 10^{-3} \; \text{eV}^2 $$
$$ \text{sin}^2\theta_{12}=0.27-0.34 $$
$$ \text{sin}^2\theta_{23}=0.34-0.67 $$ 
\begin{equation}
\text{sin}^2\theta_{13}=0.016-0.030
\label{equation:data1}
\end{equation}
Another global fit study \cite{fogli} reports the 3$\sigma$ values as
$$ \Delta m_{21}^2=(6.99-8.18) \times 10^{-5} \; \text{eV}^2$$
$$ \Delta m_{31}^2 \;(\text{NH}) =(2.19-2.62)\times 10^{-3} \; \text{eV}^2 $$
$$ \Delta m_{23}^2 \;(\text{IH}) =(2.17-2.61)\times 10^{-3} \; \text{eV}^2 $$
$$ \text{sin}^2\theta_{12}=0.259-0.359 $$
$$ \text{sin}^2\theta_{23}=0.331-0.637 $$ 
\begin{equation}
\text{sin}^2\theta_{13}=0.017-0.031
\label{equation:data2}
\end{equation}
where NH and IH refers to normal and inverted hierarchies respectively. Although the $3\sigma$ range for the Dirac CP phase is $0-2\pi$, there are two possible best fit values of it found in the literature: $5\pi/3$ \cite{schwetz12} and $\pi$ \cite{fogli}.

The above recent data have clear evidence for non-zero $\theta_{13}$, which was earlier considered to be zero or negligibly small. This has led to a great deal of activities in neutrino model building \cite{nzt13, nzt13GA}. These frameworks which predict non-zero $\theta_{13}$ may also shed light on the Dirac CP violating phase which is still unknown. Apart from the Dirac CP phase, the nature of neutrino mass hierarchy is also an important yet unresolved issue. Understanding the correct nature of hierarchy can also have important relevance in the production of matter antimatter asymmetry (or baryon asymmetry) in the early Universe. The observed baryon asymmetry in the Universe is encoded in the baryon to photon ratio measured by dedicated cosmology experiments like Wilkinson Mass Anisotropy Probe (WMAP), Planck etc. The baryon to photon ratio in nine year WMAP data \cite{wmap9} is found to be  
\begin{equation}
Y_B \simeq (6.19 \pm 0.14) \times 10^{-10}
\end{equation} 
The latest data available from Planck mission constrain the baryon to photon ratio \cite{Planck13} as
\begin{equation}
Y_B \simeq (6.065 \pm 0.090) \times 10^{-10}
\label{barasym}
\end{equation} 
Leptogenesis is one of the most widely studied mechanism to generate this observed baryon asymmetry in the Universe by generating an asymmetry in the lepton sector first which is later converted into baryon asymmetry through electroweak sphaleron transitions \cite{sphaleron}. As pointed out by Fukugita and Yanagida \cite{fukuyana}, the out of equilibrium and CP violating decay of heavy Majorana neutrinos provides a natural way to create the required lepton asymmetry. The novel feature of this mechanism is the way it relates two of the most widely studied problems in particle physics: the origin of neutrino mass and the origin of matter-antimatter asymmetry. This idea has been implemented in several interesting models in the literature \cite{leptoreview,joshipura}. Apart from providing a solution to the matter-antimatter asymmetry problem, the study of leptogenesis can also put some indirect constraints on the CP phases and neutrino hierarchies which are yet unsettled at neutrino oscillation experiments.

In view of above, the present work is planned to carry out a study of baryogenesis through leptogenesis in neutrino mass models with normal and inverted hierarchical neutrino masses within the framework of generic left-right symmetric models (LRSM) \cite{lrsm}. Such a work was done recently in \cite{mkd-db-rm} where the viability of several scenarios were studied by fitting their predictions to the best fit neutrino oscillation data in the presence of both type I and type II seesaw. In this present work, we scan the parameter space of the same sets of models by calculating the baryon asymmetry and comparing with the experimental data. We first parametrize the neutrino mass matrix using central values of the global fit neutrino oscillation data. We then keep the Majorana CP phases as free parameters and find out the allowed ranges for which correct baryon asymmetry can be generated. This calculation is performed first by considering type I seesaw and taking the Dirac neutrino mass matrix in a diagonal form. First we choose the Dirac neutrino mass matrix in such a way that the lightest right handed neutrino mass is $M_1 > 10^{12}$ GeV so as to ignore the lepton flavor effects. Keeping in view of the fact that the lepton flavor effects are very important in leptogenesis as studied initially by the authors in \cite{flavorlepto} and later explored in great details in many other works in the literature, we consider the flavor effects in the subsequent analysis.  For that, we choose the Dirac neutrino mass matrix in such a way that the lightest right handed neutrino mass falls in the range appropriate for the particular flavor effects to be present. 

After calculating baryon asymmetry for type I seesaw case, we then consider the contribution of both type I and type II seesaw terms to the neutrino mass with type I term dominating and type II term as a small perturbation. The strength of the type II seesaw term is kept at its minimum for a particular choice of left-right symmetry breaking scale and the prediction for baryon asymmetry is calculated by varying the Majorana CP phases. We note from our analysis that certain combinations of CP phases and lightest neutrino mass eigenvalue are disfavored from the demand of successful thermal leptogenesis. This can certainly put some indirect bounds on these free parameters which are unconstrained from present neutrino experiments.

This paper is organized as follows: in section \ref{method} we discuss the methodology of type I and type II seesaw mechanism in generic LRSM. In section \ref{lepto}, we outline the details of computing baryon asymmetry in LRSM. In section \ref{numeric} we discuss our numerical analysis and results and then finally conclude in section \ref{conclude}.
\section{Seesaw in LRSM}
\label{method}
Type I seesaw framework is the simplest possible seesaw mechanism  and can arise in simple extensions of the standard model by three right handed neutrinos. There is also another type of non-canonical seesaw (known as type-II seesaw)\cite{tii} where  a left-handed Higgs triplet $\Delta_{L}$ picks up a vacuum expectation value (vev). This is possible both in the minimal extension of the standard model by $\Delta_{L}$ or in other well motivated extensions like left-right symmetric models (LRSM) \cite{lrsm}. The seesaw formula in LRSM can be written as
\begin{equation}
m_{LL}=m_{LL}^{II} + m_{LL}^I
\label{type2a}
\end{equation}
 where the usual type I seesaw formula is given by the expression,
\begin{equation}
m_{LL}^I=-m_{LR}M_{RR}^{-1}m_{LR}^{T}.
\end{equation}
Here  $m_{LR}$ is the Dirac neutrino mass matrix. The above seesaw formula with both type I and type II contributions can naturally arise in extension of standard model with three right handed neutrinos and one copy of $\Delta_{L}$. However, we will use this formula in the framework of LRSM where $M_{RR}$ arises naturally as a result of parity breaking at high energy and both the type I and type II terms can be written in terms of $M_{RR}$. In LRSM with Higgs triplets, $M_{RR}$ can be expressed as $M_{RR}=v_{R}f_{R}$ with $v_{R}$ being the vev of the right handed triplet Higgs field $\Delta_R$ imparting Majorana masses to the right-handed neutrinos and $f_{R}$ is the corresponding Yukawa coupling. The first term $m_{LL}^{II}$ in equation (\ref{type2a}) is due to the vev of $SU(2)_{L}$ Higgs triplet. Thus, $m_{LL}^{II}=f_{L}v_{L}$ and $M_{RR}=f_{R}v_{R}$, where $v_{L,R}$ denote the vev's and $f_{L,R}$ are symmetric $3\times3$ matrices. The left-right symmetry demands $f_{R}=f_{L}=f$. The induced vev for the left-handed triplet $v_{L}$ can be shown for generic LRSM to be
$$v_{L}=\gamma \frac{M^{2}_{W}}{v_{R}}$$
with $M_{W}\sim 80.4$ GeV being the weak boson mass such that 
$$ |v_{L}|<<M_{W}<<|v_{R}| $$ 
In general $\gamma$ is a function of various couplings in the scalar potential of generic LRSM and without any fine tuning $\gamma$ is expected to be of the order unity ($\gamma\sim 1$). type-II seesaw formula in equation (\ref{type2a}) can now be expressed as
\begin{equation}
m_{LL}=\gamma (M_{W}/v_{R})^{2}M_{RR}-m_{LR}M^{-1}_{RR}m^{T}_{LR}
\label{type2}
\end{equation}

With above seesaw formula (\ref{type2}), the neutrino mass matrices
are constructed by considering contributions from both type I and type II terms. The choice of $v_R$ however, remains ambiguous in the literature where different choices of $v_{R}$ are made according to convenience \cite{bstn,bd,ca,aw,joshipura}. However, in this present work we will always take $v_{R}$ as $v_R=\gamma\frac{M^2_W}{v_L} \simeq \gamma \times 10^{15}\;\text{GeV}$ \cite{aw, joshipura}. It is worth mentioning that, here $SU(2)_R \times U(1)_{B-L}$ gauge symmetry breaking scale (as in generic LRSM) $v_R$ is the same as the scale of parity breaking \cite{bstn}. Using this form of $v_R$, the seesaw formula (\ref{type2}) becomes 
\begin{equation}
m_{LL}=\gamma \left (\frac{M_{W}}{\gamma \times 10^{15}} \right )^{2}M_{RR}-m_{LR}M^{-1}_{RR}m^{T}_{LR}
\label{type2b}
\end{equation}

Quantitatively, either of the two terms on the right hand side of equation (\ref{type2b}) can be dominant or both the terms can be equally dominant. However, for generic choices of symmetry breaking scales (mentioned above) as well as the Dirac neutrino mass matrices (generically to be of same order as corresponding charged lepton masses), both type I and type II term can be equally dominant only when the dimensionless parameter $\gamma$ is fine tuned to be very small. We check this by equating both the terms to the best fit value of $m_{LL}$. We skip such a fine-tuned case here and consider the case in which type I term dominates whereas type II term is present as a small perturbation.

In this case, the second term in the equation (\ref{type2b}) gives the leading contribution to $m_{LL}$ and hence we compute the right-handed neutrino mass matrix $M_{RR}$ by using the inverse type I seesaw formula $M_{RR}=m_{LR}^Tm_{LL}^{-1}m_{LR}$ where we use the best fit $m_{LL}$ and generic $m_{LR}$ as will be shown in the section \ref{numeric}. Here we hold $M_{RR}$ fixed, so the first term in equation (\ref{type2b}) is dependent on the value of $\gamma$ while second term is fixed. For $\gamma \sim 1$, the first term has minimum contribution whereas for smaller values of $\gamma$, the contribution of the type II term will increase. We fix the dimensionless parameter $\gamma$ at $1.0$ such that the type II contribution is minimal. Taking a larger contribution of type II seesaw term will disturb the neutrino oscillation parameters from their best fit values since we have used the best fit neutrino parameters for type I seesaw.

\begin{figure}
\begin{center}
\includegraphics[width=0.5\textwidth]{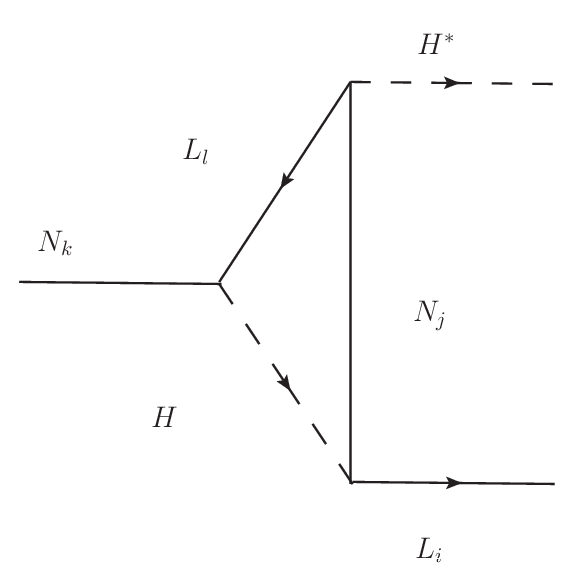}
\end{center}
\caption{Right handed neutrino decay}
\label{fig0001}
\end{figure}
\begin{figure}
\begin{center}
\includegraphics[width=0.5\textwidth]{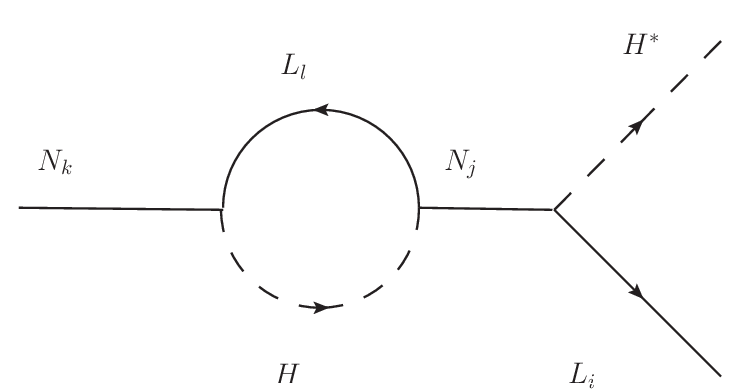}
\end{center}
\caption{Right handed neutrino decay}
\label{fig001}
\end{figure}
\begin{figure}
\begin{center}
\includegraphics[width=0.5\textwidth]{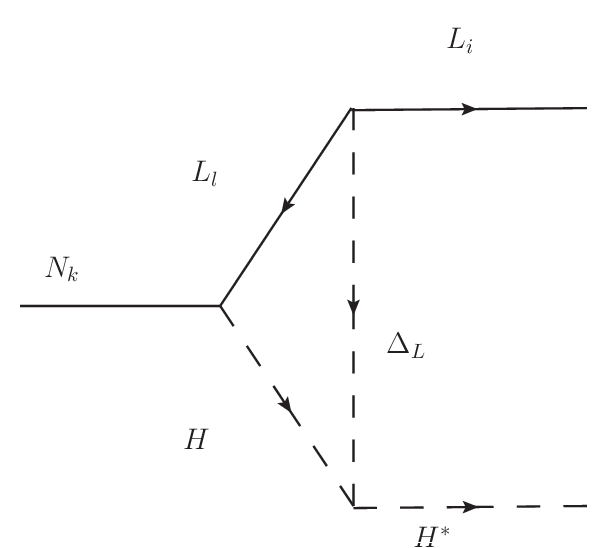}
\end{center}
\caption{Right handed neutrino decay}
\label{fig01}
\end{figure}

\begin{figure}[h]
\begin{center}
$
\begin{array}{cc}
\includegraphics[width=0.5\textwidth]{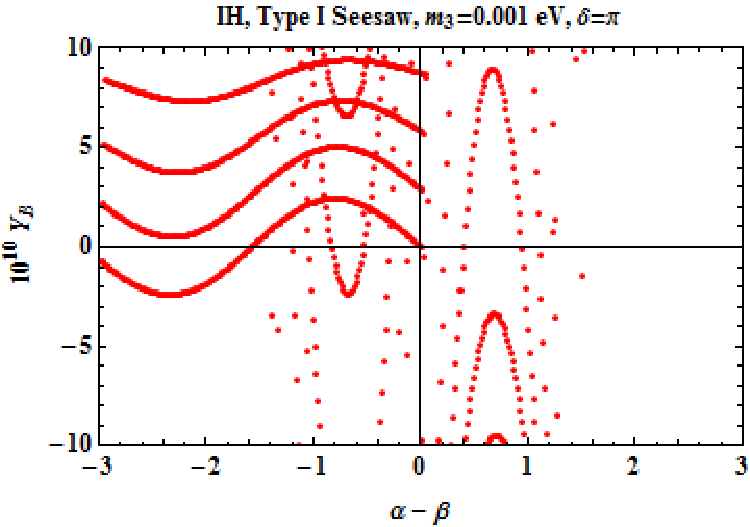} &
\includegraphics[width=0.5\textwidth]{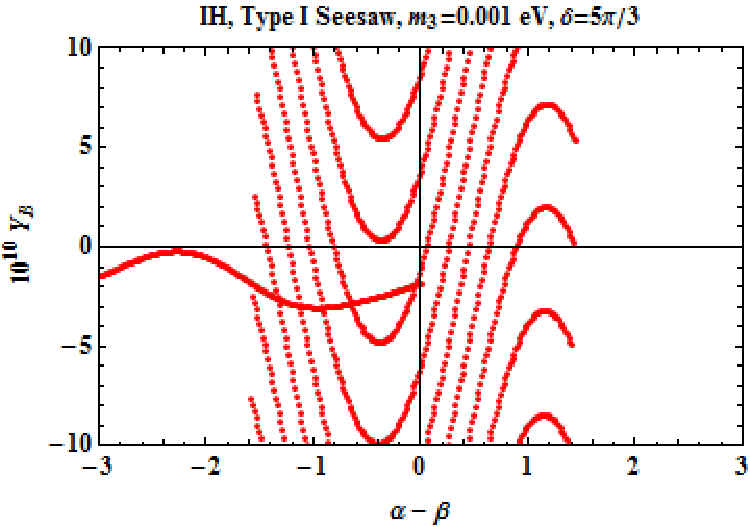} \\
\includegraphics[width=0.5\textwidth]{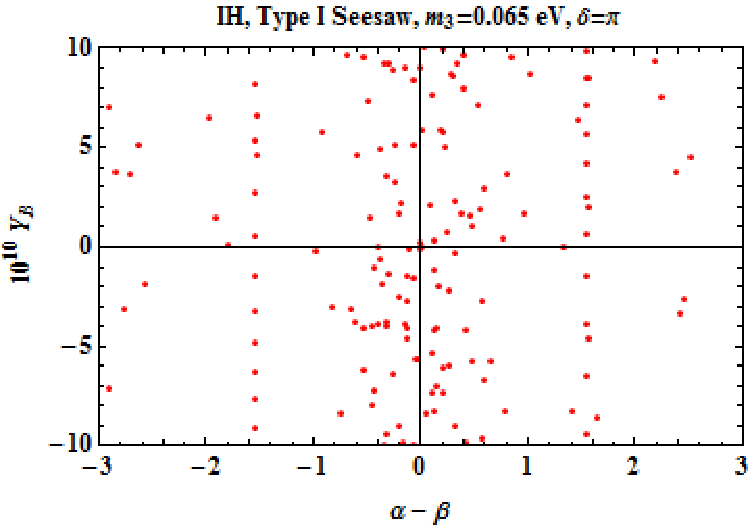} &
\includegraphics[width=0.5\textwidth]{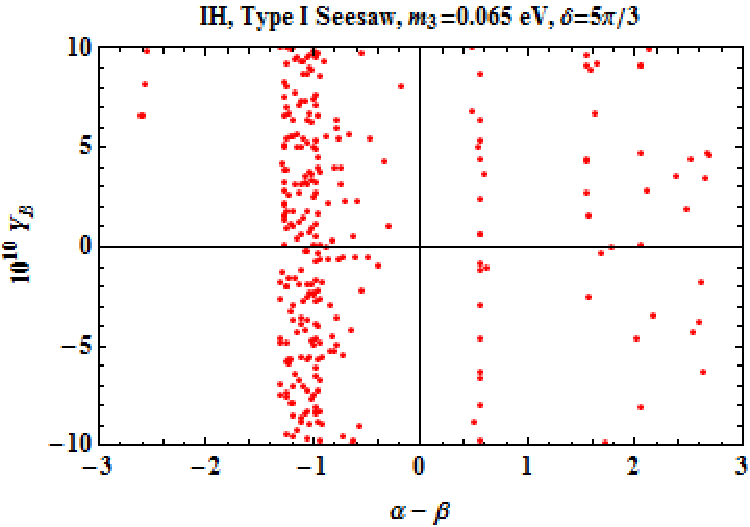}
\end{array}$
\end{center}
\caption{Predictions for baryon to photon ratio as a function of the difference in Majorana CP phases with type I seesaw for inverted hierarchy in the one flavor regime}
\label{fig1}
\end{figure}
\begin{figure}[h]
\begin{center}
$
\begin{array}{cc}
\includegraphics[width=0.5\textwidth]{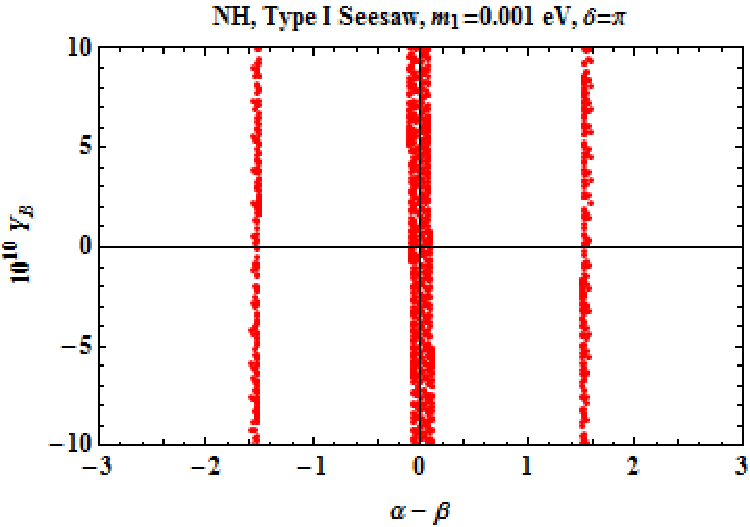} &
\includegraphics[width=0.5\textwidth]{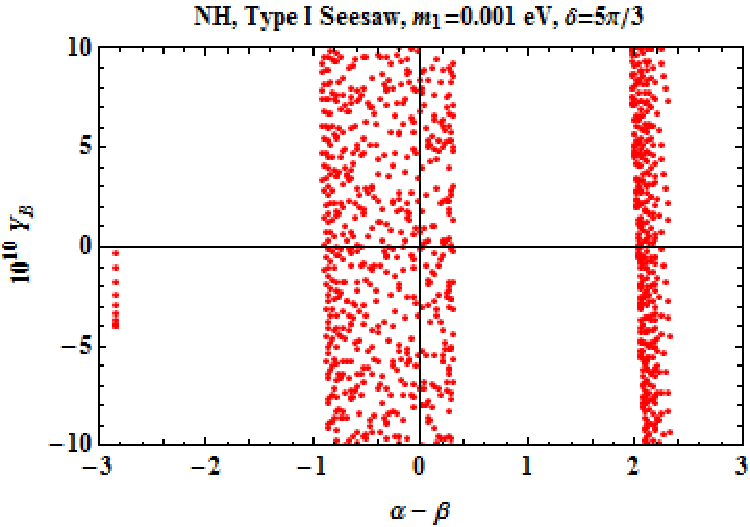} \\
\includegraphics[width=0.5\textwidth]{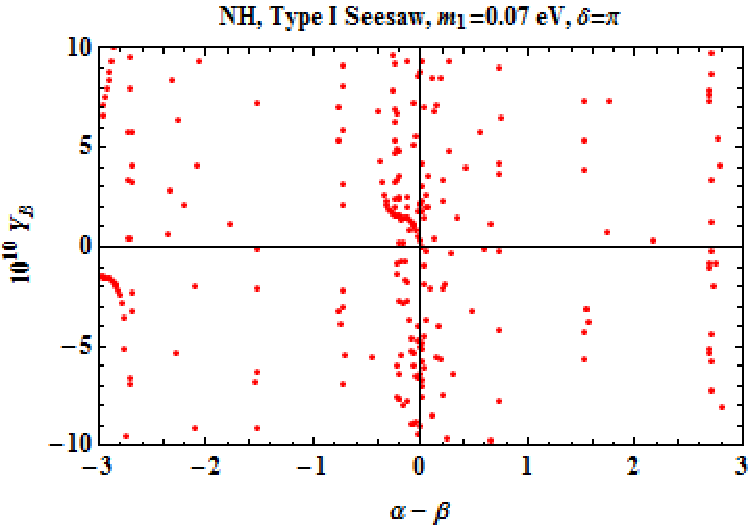} &
\includegraphics[width=0.5\textwidth]{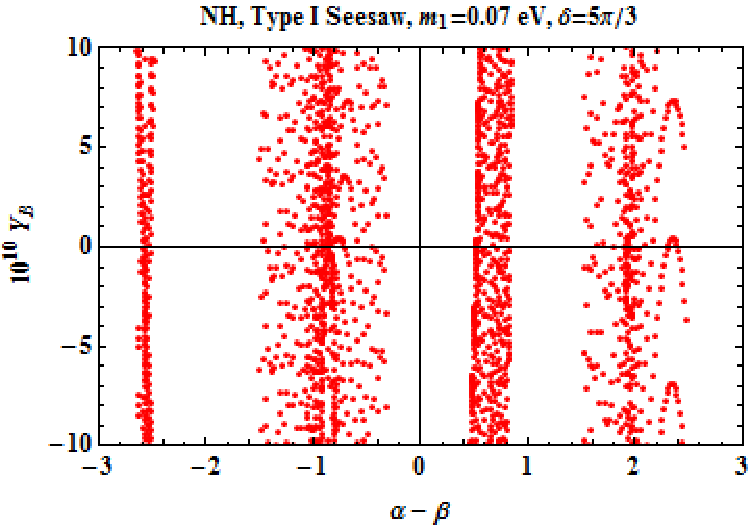}
\end{array}$
\end{center}
\caption{Predictions for baryon to photon ratio as a function of the difference in Majorana CP phases with type I seesaw for normal hierarchy in the one flavor regime}
\label{fig2}
\end{figure}
\begin{figure}[h]
\begin{center}
$
\begin{array}{cc}
\includegraphics[width=0.5\textwidth]{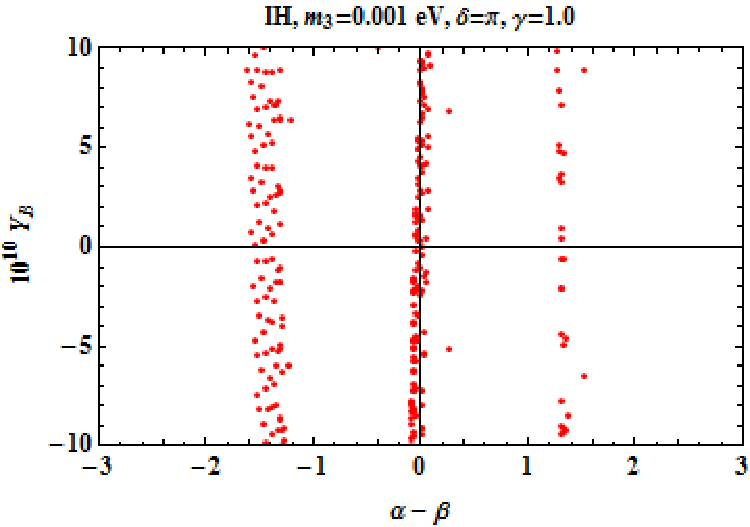} &
\includegraphics[width=0.5\textwidth]{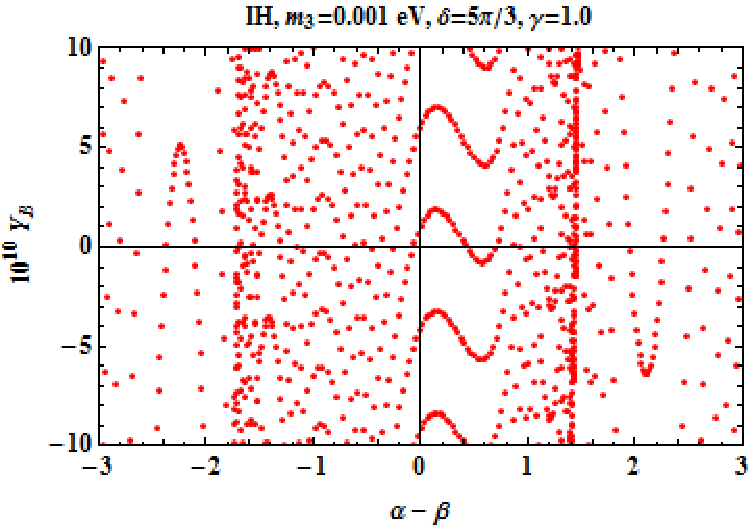} \\
\includegraphics[width=0.5\textwidth]{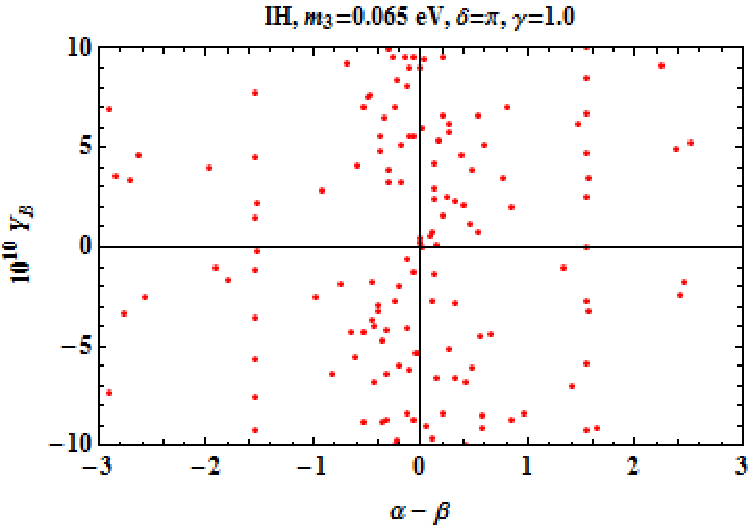} &
\includegraphics[width=0.5\textwidth]{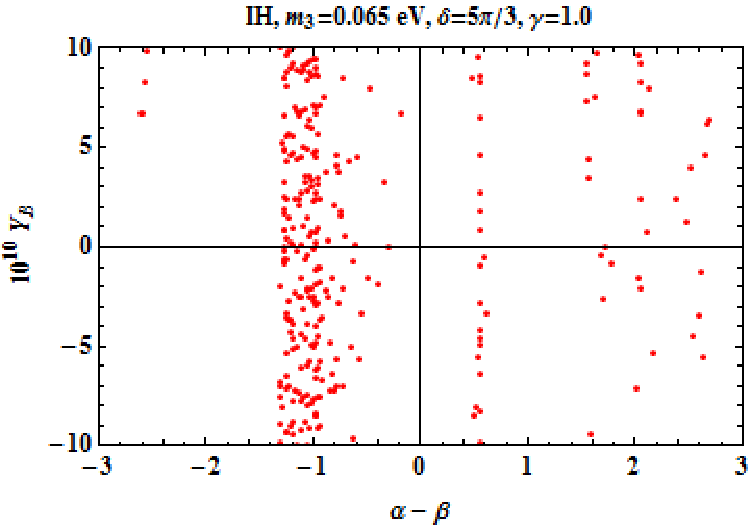}
\end{array}$
\end{center}
\caption{Predictions for baryon to photon ratio as a function of the difference in Majorana CP phases with type I+II seesaw for inverted hierarchy in the one flavor regime}
\label{fig3}
\end{figure}

\begin{figure}[h]
\begin{center}
$
\begin{array}{cc}
\includegraphics[width=0.5\textwidth]{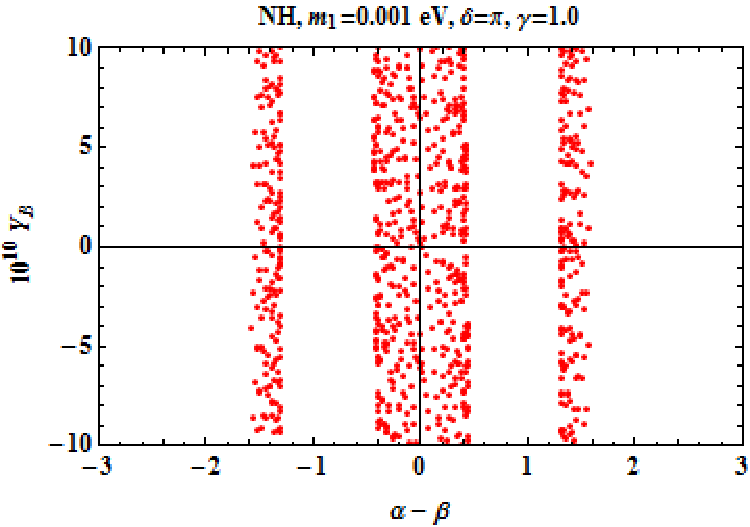} &
\includegraphics[width=0.5\textwidth]{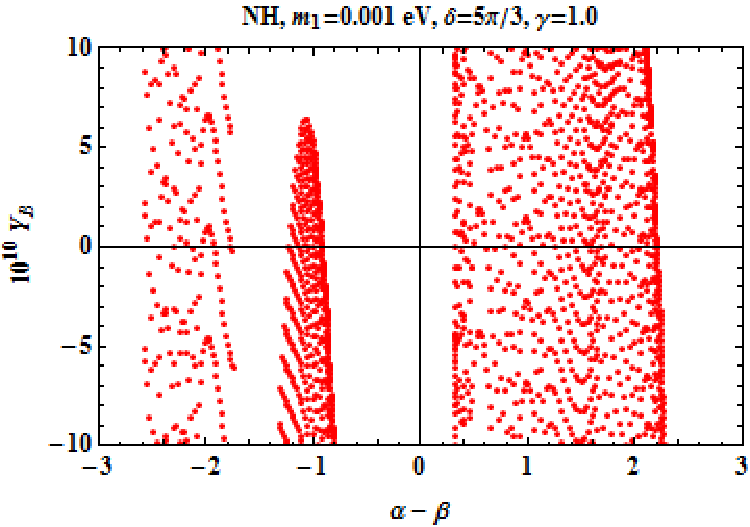} \\
\includegraphics[width=0.5\textwidth]{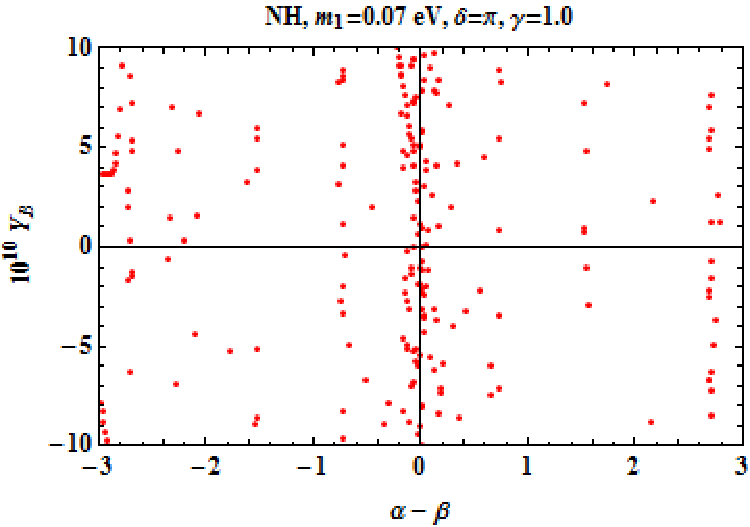} &
\includegraphics[width=0.5\textwidth]{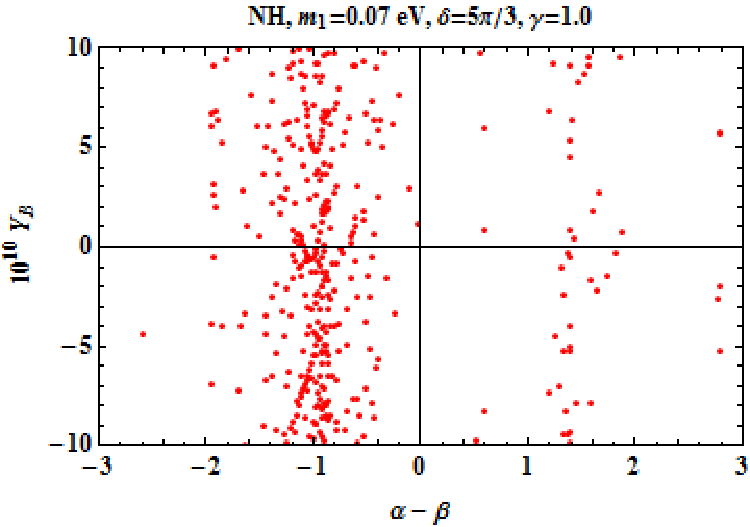}
\end{array}$
\end{center}
\caption{Predictions for baryon to photon ratio as a function of the difference in Majorana CP phases with type I+II seesaw for normal hierarchy in the one flavor regime}
\label{fig4}
\end{figure}

\begin{figure}[h]
\begin{center}
$
\begin{array}{cc}
\includegraphics[width=0.5\textwidth]{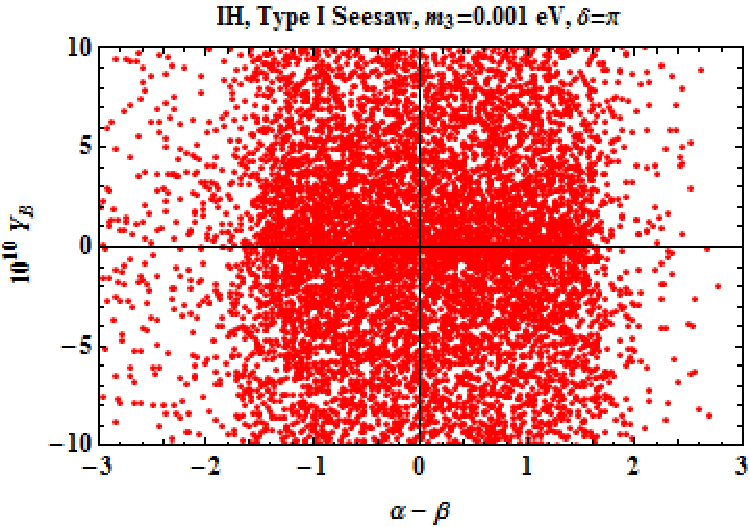} &
\includegraphics[width=0.5\textwidth]{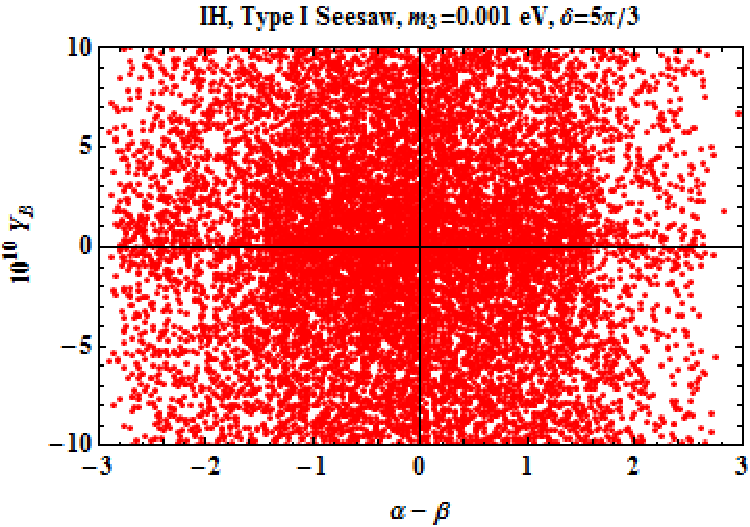} \\
\includegraphics[width=0.5\textwidth]{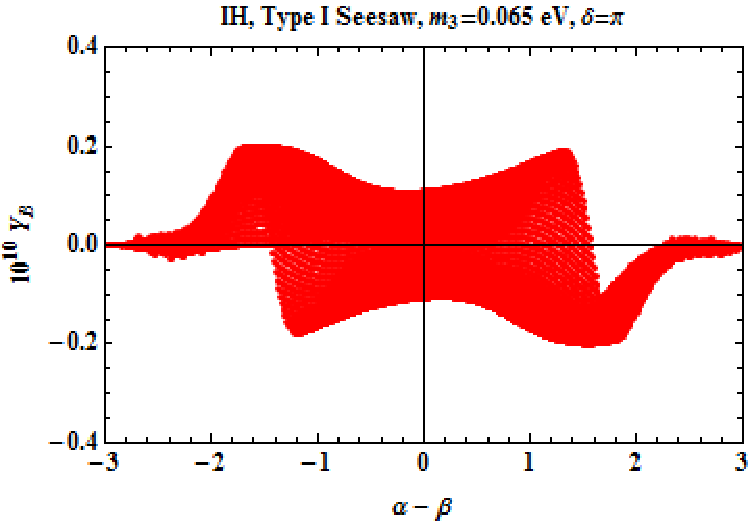} &
\includegraphics[width=0.5\textwidth]{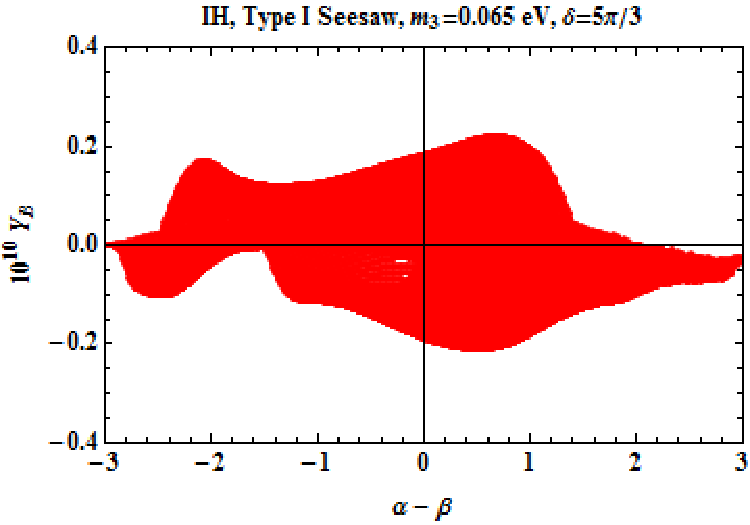}
\end{array}$
\end{center}
\caption{Predictions for baryon to photon ratio as a function of the difference in Majorana CP phases with type I seesaw for inverted hierarchy in the two flavor regime}
\label{fig5}
\end{figure}
\begin{figure}[h]
\begin{center}
$
\begin{array}{cc}
\includegraphics[width=0.5\textwidth]{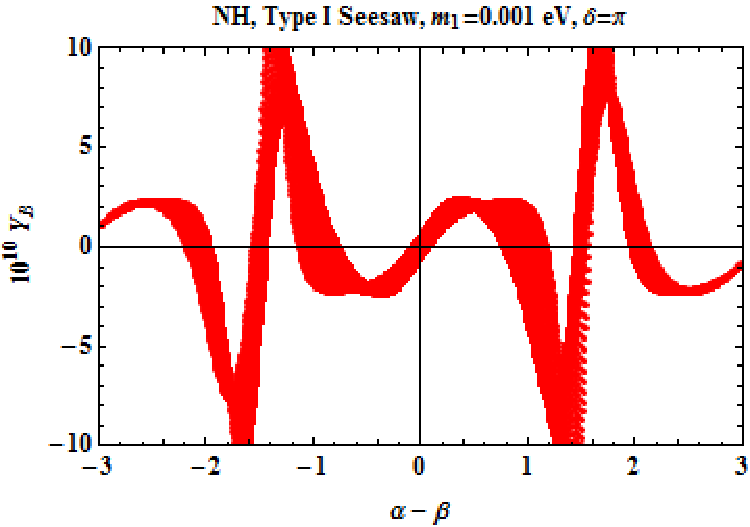} &
\includegraphics[width=0.5\textwidth]{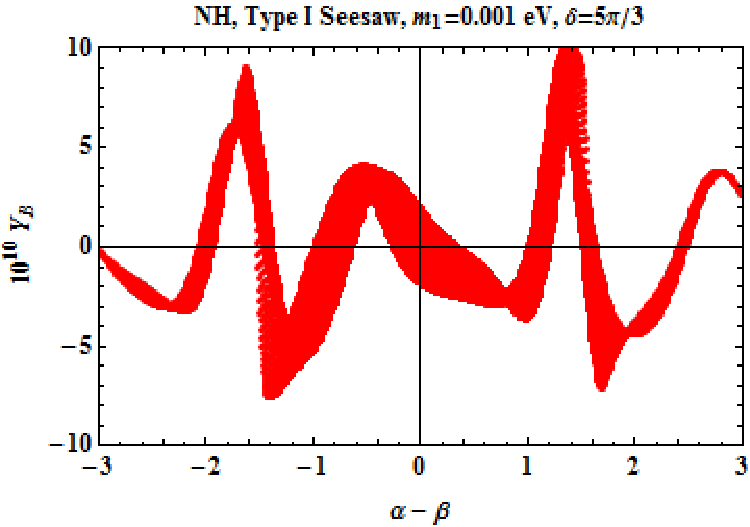} \\
\includegraphics[width=0.5\textwidth]{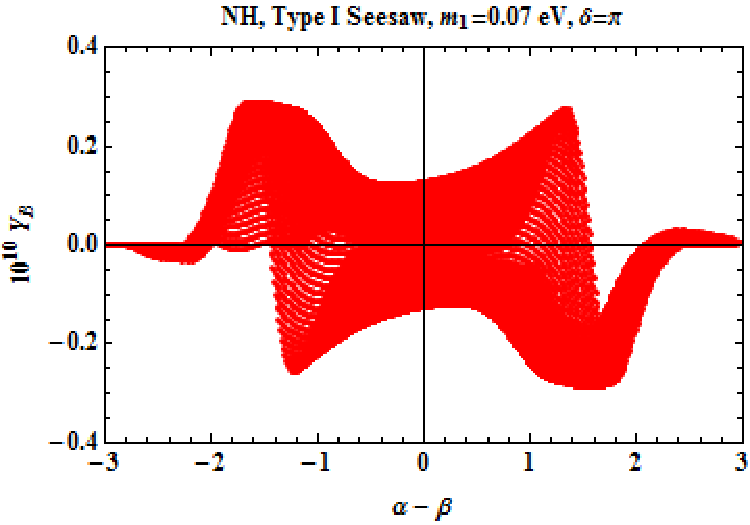} &
\includegraphics[width=0.5\textwidth]{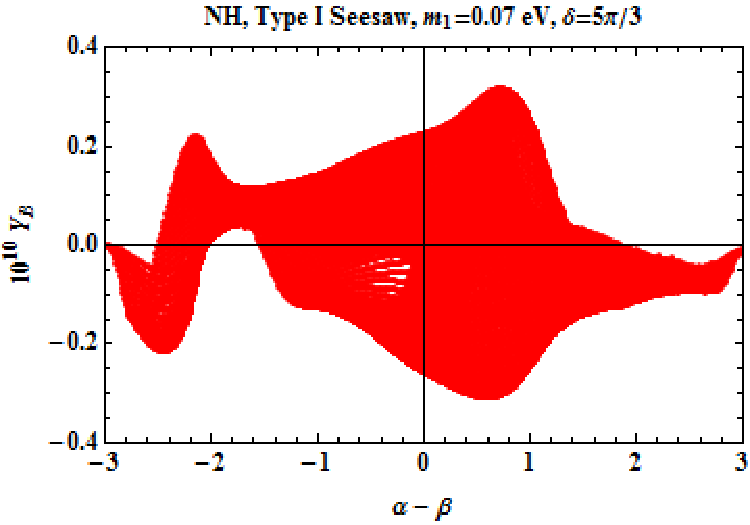}
\end{array}$
\end{center}
\caption{Predictions for baryon to photon ratio as a function of the difference in Majorana CP phases with type I seesaw for normal hierarchy in the two flavor regime}
\label{fig6}
\end{figure}
\begin{figure}[h]
\begin{center}
$
\begin{array}{cc}
\includegraphics[width=0.5\textwidth]{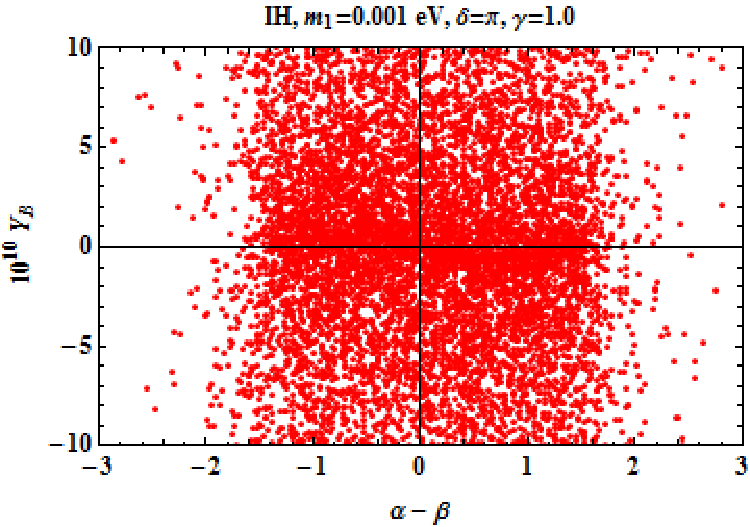} &
\includegraphics[width=0.5\textwidth]{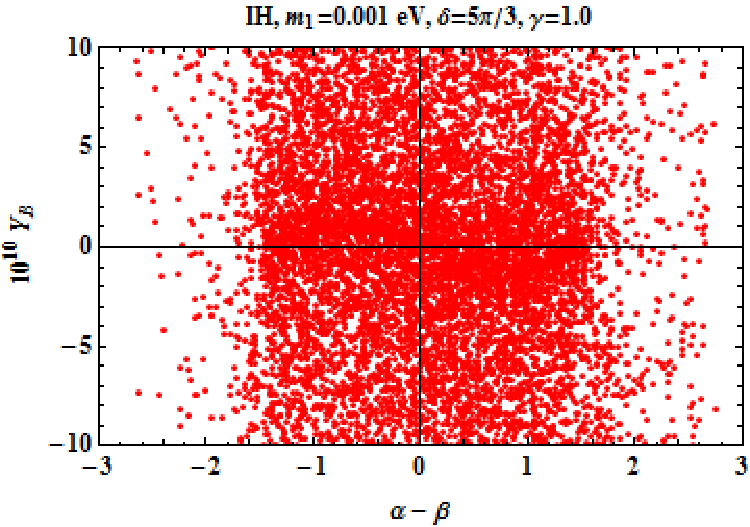} \\
\includegraphics[width=0.5\textwidth]{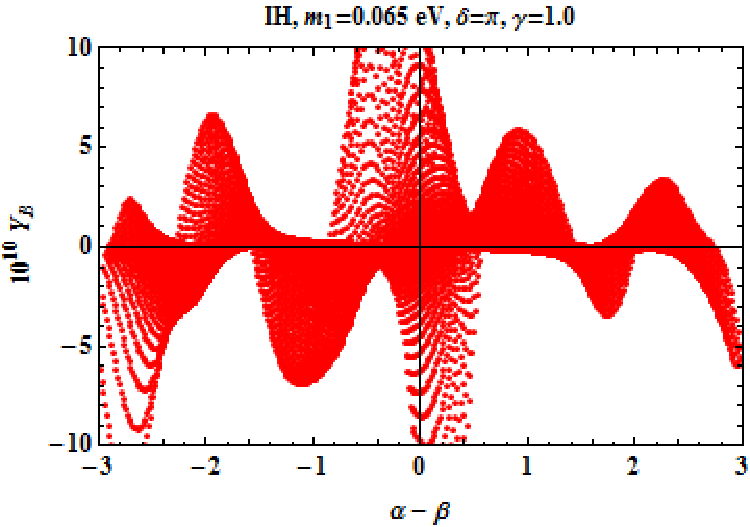} &
\includegraphics[width=0.5\textwidth]{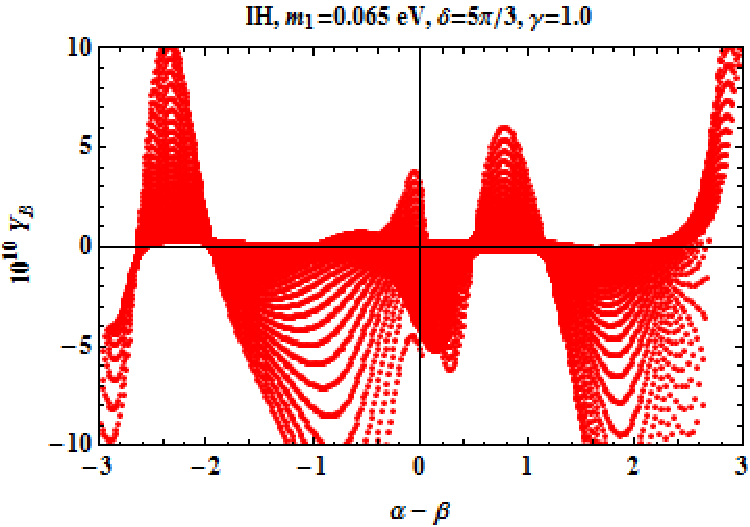}
\end{array}$
\end{center}
\caption{Predictions for baryon to photon ratio as a function of the difference in Majorana CP phases with type I+II seesaw for inverted hierarchy in the two flavor regime}
\label{fig7}
\end{figure}
\begin{figure}[h]
\begin{center}
$
\begin{array}{cc}
\includegraphics[width=0.5\textwidth]{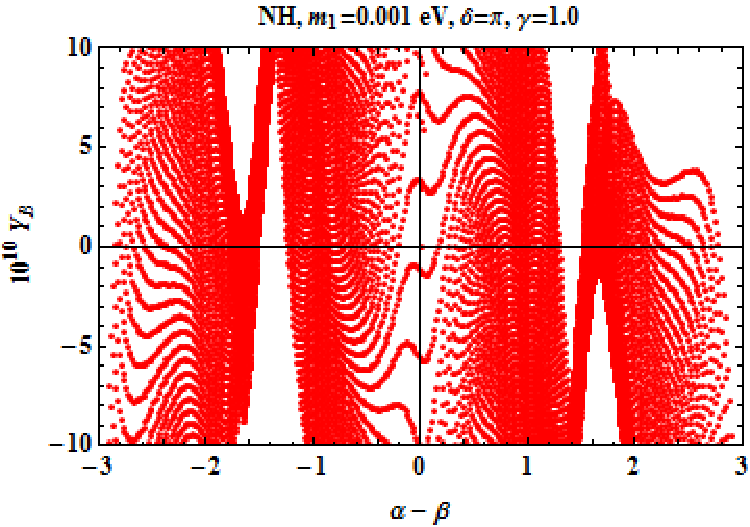} &
\includegraphics[width=0.5\textwidth]{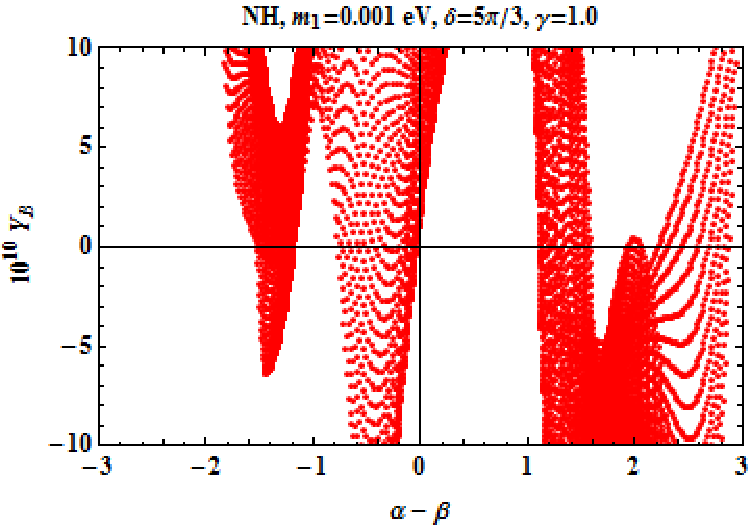} \\
\includegraphics[width=0.5\textwidth]{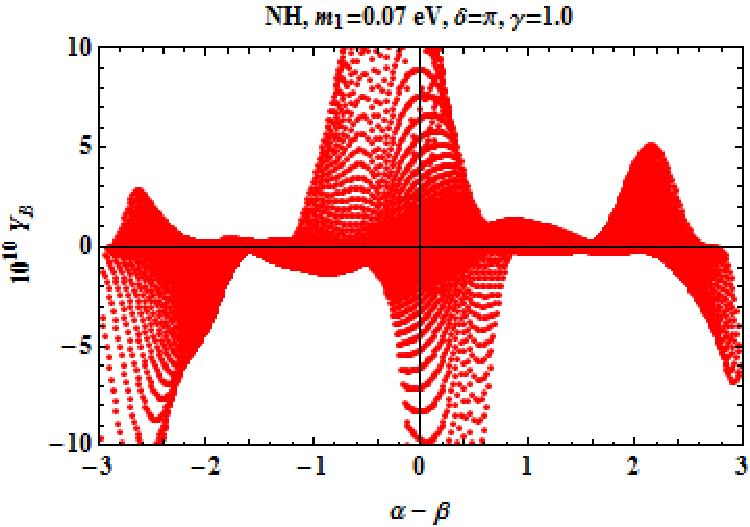} &
\includegraphics[width=0.5\textwidth]{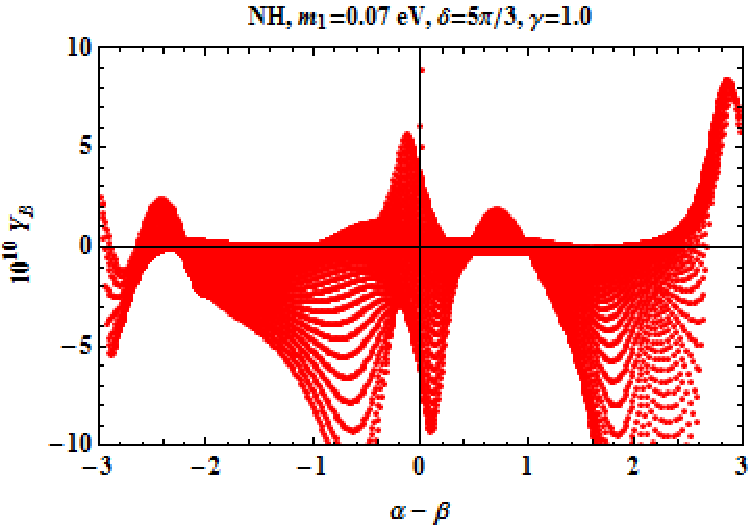}
\end{array}$
\end{center}
\caption{Predictions for baryon to photon ratio as a function of the difference in Majorana CP phases with type I+II seesaw for normal hierarchy in the two flavor regime}
\label{fig8}
\end{figure}
\begin{figure}[h]
\begin{center}
$
\begin{array}{cc}
\includegraphics[width=0.5\textwidth]{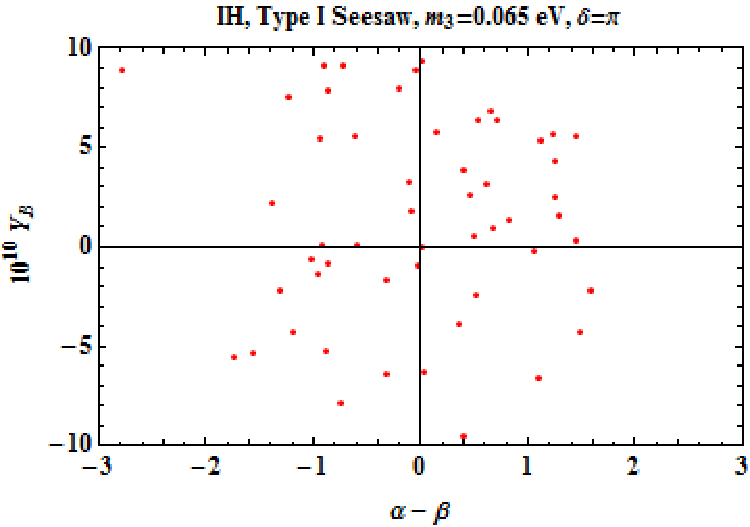} &
\includegraphics[width=0.5\textwidth]{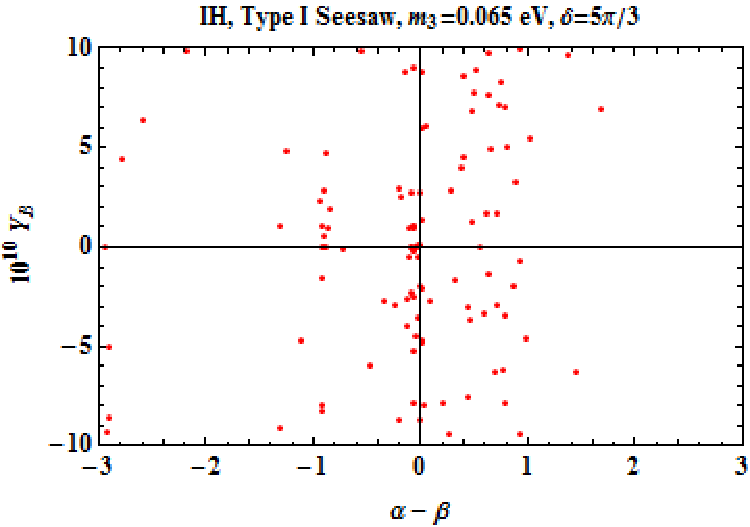}
\end{array}$
\end{center}
\caption{Predictions for baryon to photon ratio as a function of the difference in Majorana CP phases with type I seesaw for inverted hierarchy in the three flavor regime}
\label{fig9}
\end{figure}
\begin{figure}[h]
\begin{center}
$
\begin{array}{cc}
\includegraphics[width=0.5\textwidth]{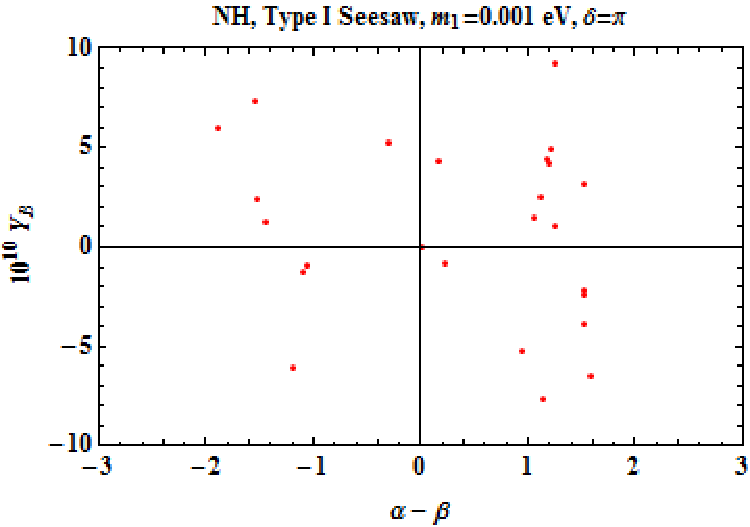} &
\includegraphics[width=0.5\textwidth]{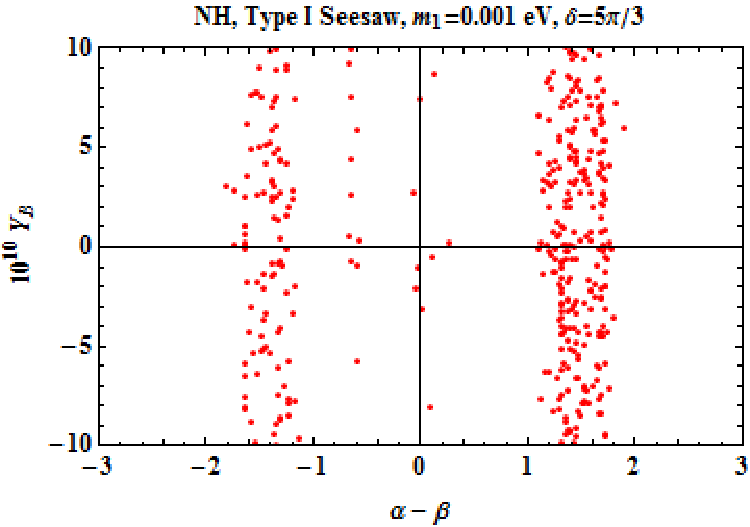} \\
\includegraphics[width=0.5\textwidth]{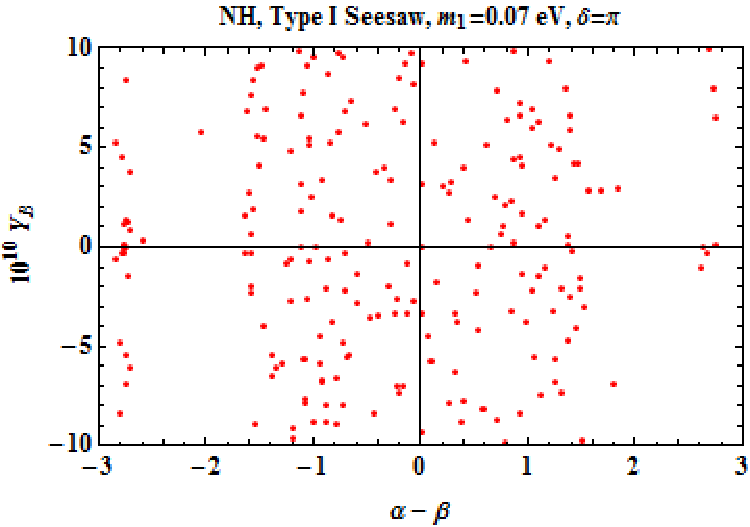} &
\includegraphics[width=0.5\textwidth]{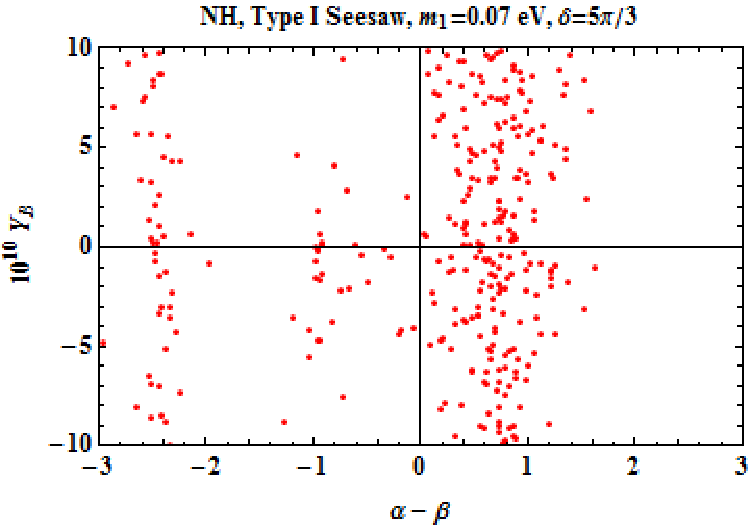}
\end{array}$
\end{center}
\caption{Predictions for baryon to photon ratio as a function of the difference in Majorana CP phases with type I seesaw for normal hierarchy in the three flavor regime}
\label{fig10}
\end{figure}
\begin{figure}[h]
\begin{center}
$
\begin{array}{cc}
\includegraphics[width=0.5\textwidth]{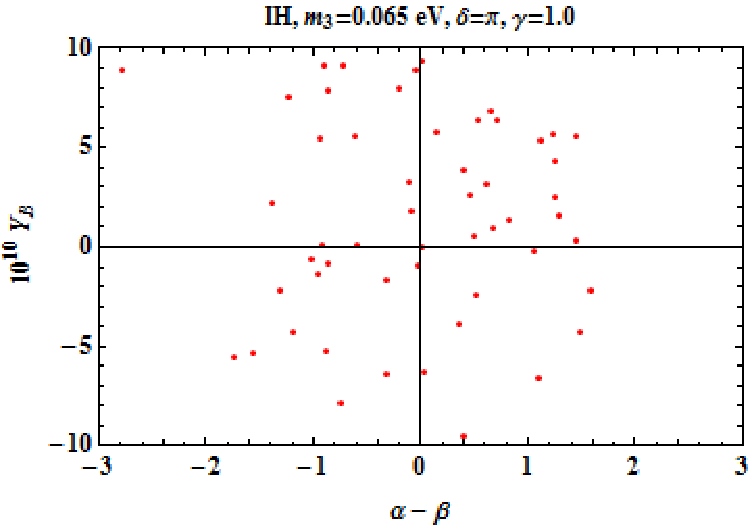} &
\includegraphics[width=0.5\textwidth]{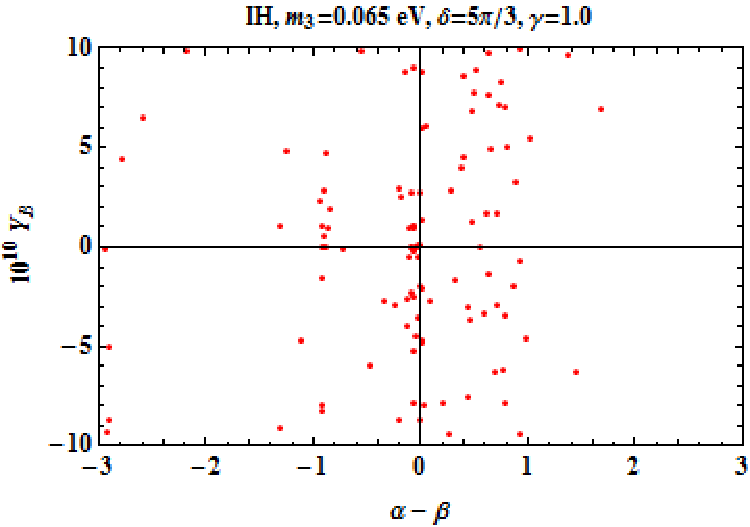}
\end{array}$
\end{center}
\caption{Predictions for baryon to photon ratio as a function of the difference in Majorana CP phases with type I+II seesaw for inverted hierarchy in the three flavor regime}
\label{fig11}
\end{figure}
\begin{figure}[h]
\begin{center}
$
\begin{array}{cc}
\includegraphics[width=0.5\textwidth]{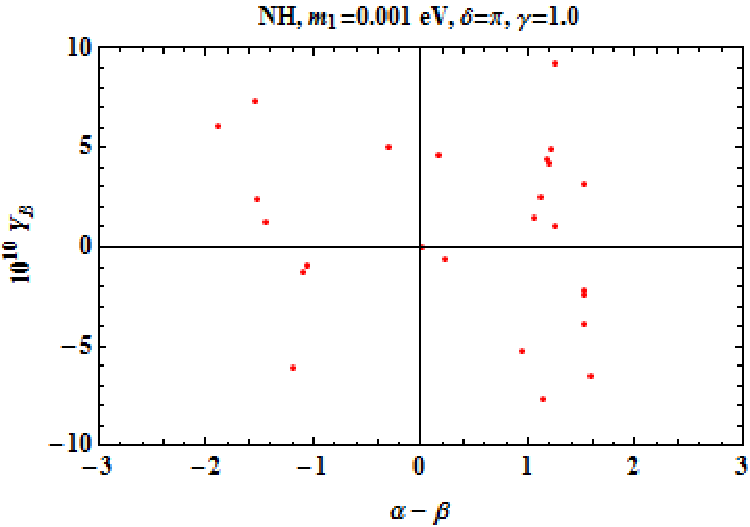} &
\includegraphics[width=0.5\textwidth]{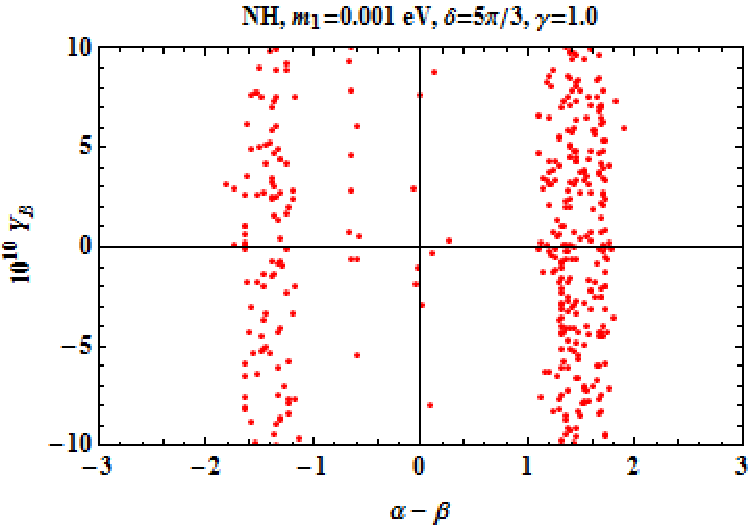} \\
\includegraphics[width=0.5\textwidth]{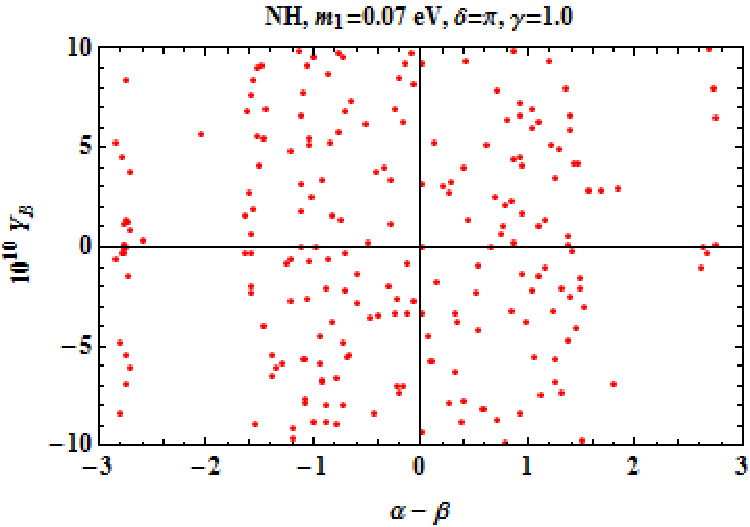} &
\includegraphics[width=0.5\textwidth]{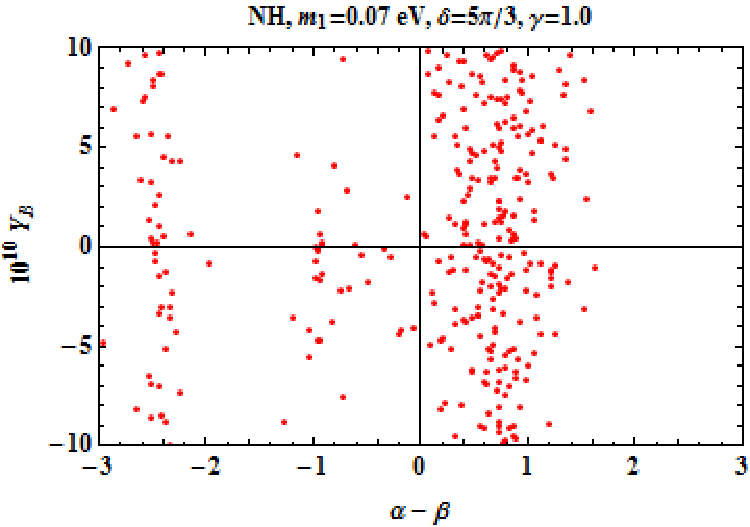}
\end{array}$
\end{center}
\caption{Predictions for baryon to photon ratio as a function of the difference in Majorana CP phases with type I+II seesaw for normal hierarchy in the three flavor regime}
\label{fig12}
\end{figure}
\section{Leptogenesis in LRSM}
\label{lepto}
As stressed earlier, leptogenesis is a novel mechanism to account for the baryon asymmetry of the Universe by creating an asymmetry in the leptonic sector first, which subsequently gets converted into baryon asymmetry through $B+L$ violating sphaleron processes during electroweak phase transition. Since quark sector CP violation is not sufficient for producing the observed baryon asymmetry, a framework explaining non-zero $\theta_{13}$ and leptonic CP phase could not only give a better picture of leptonic flavor structure, but also the origin of matter-antimatter asymmetry. 

In a model with both type I and type II seesaw mechanisms at work, there are two possible sources of lepton asymmetry: either the CP violating decay of right handed neutrino or that of scalar triplet. In our work we are considering dominant type I and sub-dominant type II seesaw which naturally point towards heavier triplet than right handed neutrinos. For simplicity we consider only the right handed neutrino decay as a source of lepton asymmetry and neglect the contribution coming from triplet decay. The lepton asymmetry from the decay of right handed neutrino into leptons and Higgs scalar is given by
\begin{equation}
\epsilon_{N_k} = \sum_i \frac{\Gamma(N_k \rightarrow L_i +H^*)-\Gamma (N_k \rightarrow \bar{L_i}+H)}{\Gamma(N_k \rightarrow L_i +H^*)+\Gamma (N_k \rightarrow \bar{L_i}+H)}
\end{equation}
In a hierarchical pattern for right handed neutrinos $M_{2,3} \gg M_1$, it is sufficient to consider the lepton asymmetry produced by the decay of the lightest right handed neutrino $N_1$. In a type I seesaw framework where the particle content is just the standard model with three additional right handed neutrinos, the lepton asymmetry is generated through the decay processes shown in figure \ref{fig0001} and \ref{fig001}. In the presence of type II seesaw, $N_1$ can also decay through a virtual triplet as can be seen in figure \ref{fig01}. Following the notations of \cite{joshipura}, the lepton asymmetry arising from the decay of $N_1$ in the presence of type I seesaw only can be written as
\begin{eqnarray}
\epsilon &=& \sum_i \frac{\Gamma(N_k \rightarrow L_i +H^*)-\Gamma (N_k \rightarrow \bar{L_i}+H)}{\Gamma(N_k \rightarrow L_i +H^*)+\Gamma (N_k \rightarrow \bar{L_i}+H)} \nonumber \\
 &=& \frac{1}{8\pi v^2}\frac{1}{(m^{\dagger}_{LR}m_{LR})_{11}} \sum_{j = 2,3} \text{Im}(m^{\dagger}_{LR}m_{LR})^2_{1j}f(M^2_j/M^2_1)
\label{noflavor}
\end{eqnarray}
where $v = 174 \; \text{GeV}$ is the vev of the Higgs bidoublets responsible for breaking the electroweak symmetry and 
$$ f(x) = \sqrt{x} \left ( 1+\frac{1}{1-x}-(1+x)\text{ln}\frac{1+x}{x} \right) $$
which can be approximated as $f(x) = -\frac{3}{2\sqrt{x}}$ for $x \gg 1$. After determining the lepton asymmetry $\epsilon$, the corresponding baryon asymmetry can be obtained by
\begin{equation}
Y_B = c \kappa \frac{\epsilon}{g_*}
\end{equation}
through electroweak sphaleron processes \cite{sphaleron}. Here the factor $c$ is measure of the fraction of lepton asymmetry being converted into baryon asymmetry and is approximately equal to $-0.55$. $\kappa$ is the dilution factor due to wash-out process which erase the produced asymmetry and can be parametrized as \cite{kolbturner}
\begin{eqnarray}
-\kappa &\simeq &  \sqrt{0.1K} \text{exp}[-4/(3(0.1K)^{0.25})], \;\; \text{for} \; K  \ge 10^6 \nonumber \\
&\simeq & \frac{0.3}{K (\ln K)^{0.6}}, \;\; \text{for} \; 10 \le K \le 10^6 \nonumber \\
&\simeq & \frac{1}{2\sqrt{K^2+9}},  \;\; \text{for} \; 0 \le K \le 10.
\end{eqnarray}
where K is given as
$$ K = \frac{\Gamma_1}{H(T=M_1)} = \frac{(m^{\dagger}_{LR}m_{LR})M_1}{8\pi v^2} \frac{M_{Pl}}{1.66 \sqrt{g_*}M^2_1} $$
Here $\Gamma_1$ is the decay width of $N_1$ and $H(T=M_1)$ is the Hubble constant at temperature $T = M_1$. The factor $g_*$ is the effective number of relativistic degrees of freedom at $T=M_1$ and is approximately $110$. It should be noted that the above estimate for baryon asymmetry is valid only in the absence of wash-out effects responsible for erasing the asymmetry created. In left right symmetric models, right handed neutrinos also have $SU(2)_R$gauge interactions and hence can give rise to sizable wash-out effects. However, as noted in \cite{washout} such effects can be neglected if $M_1/v_R < 10^{-2}$. In the present study, we take $v_R \sim 10^{15}$ GeV as mentioned above. As we discuss in details below, the largest possible value of $M_1$ in our study is approximately $10^{13}$ GeV (one flavor regime) which just satisfies the bound $M_1/v_R < 10^{-2}$ so as to neglect the wash-out effects arising from $SU(2)_R$ gauge interactions. We also ignore the asymmetry that can be produced by the decay of the heavy Higgs triplets and consider only the right handed neutrino decay as the primary source of lepton asymmetry. 

We note that the lepton asymmetry shown in equation (\ref{noflavor}) is obtained by summing over all the flavors $\alpha = e, \mu, \tau$. A non-vanishing lepton asymmetry is generated only when the right handed neutrino decay is out of equilibrium. Otherwise both the forward and the backward processes will happen at the same rate resulting in a vanishing asymmetry. Departure from equilibrium can be estimated by comparing the interaction rate with the expansion rate of the Universe. At very high temperatures $(T \geq 10^{12} \text{GeV})$ all charged lepton flavors are out of equilibrium and hence all of them behave similarly resulting in the one flavor regime. However at temperatures $ T < 10^{12}$ GeV $(T < 10^9 \text{GeV})$, interactions involving tau (muon) Yukawa couplings enter equilibrium and flavor effects become important \cite{flavorlepto}. Taking these flavor effects into account, the final baryon asymmetry is given by 
\begin{equation}
Y^{2 flavor}_B = \frac{-12}{37g^*}[\epsilon_2 \eta\left (\frac{417}{589}\tilde{m_2} \right)+\epsilon^{\tau}_1\eta\left (\frac{390}{589}\tilde{m_{\tau}}\right )] \nonumber
\end{equation}
\begin{equation}
Y^{3 flavor}_B = \frac{-12}{37g^*}[\epsilon_e \eta\left (\frac{151}{179}\tilde{m_e}\right)+ \epsilon^{\mu}_1 \eta\left (\frac{344}{537}\tilde{m_{\mu}}\right)+\epsilon^{\tau}_1\eta\left (\frac{344}{537}\tilde{m_{\tau}} \right )] \nonumber
\end{equation}
where $\epsilon_2 = \epsilon^e_1 + \epsilon^{\mu}_1, \tilde{m_2} = \tilde{m_e}+\tilde{m_{\mu}}, \tilde{m_{\alpha}} = \frac{(m^*_{LR})_{\alpha 1} (m_{LR})_{\alpha 1}}{M_1}$ and the factor $g_*$ is the effective number of relativistic degrees of freedom at $T=M_1$ and is approximately $110$. The function $\eta$ is given by 
$$ \eta (\tilde{m_{\alpha}}) = \left [\left ( \frac{\tilde{m_{\alpha}}}{8.25 \times 10^{-3} \text{eV}} \right )^{-1}+ \left ( \frac{0.2\times 10^{-3} \text{eV}}{\tilde{m_{\alpha}}} \right )^{-1.16} \right ]^{-1} $$
In the presence of an additional scalar triplet, the right handed neutrino can also decay through a virtual triplet as shown in figure \ref{fig01}. The contribution of this diagram to lepton asymmetry can be estimated as \cite{tripletlepto}
\begin{equation} 
\epsilon^{\alpha}_{\Delta 1}=-\frac{M_1}{8\pi v^2} \frac{\sum_{j=2,3} \text{Im} [(m_{LR})_{1j}(m_{LR})_{1\alpha}(M^{II*}_{\nu})_{j\alpha}]}{\sum_{j=2,3} \lvert (m_{LR})_{1j}\rvert^2}
\label{eps3}
\end{equation}
where $M_1 \ll M_{\Delta}$ is assumed which is natural in a model with dominant type I and sub-dominant type II seesaw mechanisms.

For the calculation of baryon asymmetry, we go to the basis where the right handed Majorana neutrino mass matrix is diagonal
\begin{equation}
U^*_R M_{RR} U^{\dagger}_R = \text{diag}(M_1, M_2, M_3)
\label{mrrdiag}
\end{equation}
In this diagonal $M_{RR}$ basis, the Dirac neutrino mass matrix also changes to 
\begin{equation}
m_{LR} = m^d_{LR} U_R
\label{mlrdiag}
\end{equation}
where $m^d_{LR}$ is the assumed choice of the Dirac neutrino mass matrix in our calculation given by 
\begin{equation}
m_{LR}=\left(\begin{array}{ccc}
\lambda^m & 0 & 0\\
0 & \lambda^n & 0 \\
0 & 0 & 1
\end{array}\right)m_f
\label{mLR1}
\end{equation}
where $\lambda = 0.22$ is the standard Wolfenstein parameter and $(m,n)$ are positive integers.

\section{Numerical Analysis and Results}
\label{numeric}

For the purpose of numerical analysis we parametrize the neutrino mixing matrix as
\begin{equation}
U_L=\left(\begin{array}{ccc}
c_{12}c_{13}& s_{12}c_{13}& s_{13}e^{-i\delta}\\
-s_{12}c_{23}-c_{12}s_{23}s_{13}e^{i\delta}& c_{12}c_{23}-s_{12}s_{23}s_{13}e^{i\delta} & s_{23}c_{13} \\
s_{12}s_{23}-c_{12}c_{23}s_{13}e^{i\delta} & -c_{12}s_{23}-s_{12}c_{23}s_{13}e^{i\delta}& c_{23}c_{13}
\end{array}\right) \text{diag}(1, e^{i\alpha}, e^{i(\beta+\delta )})
\label{matrix1}
\end{equation}
where $c_{ij} = \cos{\theta_{ij}}, \; s_{ij} = \sin{\theta_{ij}}$. $\delta$ is the Dirac CP phase and $\alpha, \beta$ are the Majorana phases. Since we know only the two mass squared differences for the active neutrinos, we consider the lightest mass eigenvalue as free parameter allowing it to take two possible values. One such value is the maximum allowed by the Planck bound \cite{Planck13} on the sum of absolute neutrino masses $\sum_i |m_i| < 0.23$ eV and other value is $0.001$ eV. We take the best fit central values of the mixing angles and take two possible best fit values of the Dirac CP phase: $5\pi/3$ \cite{schwetz12} and $\pi$ \cite{fogli}.

After parameterizing the neutrino mass matrix using oscillation data, we consider type I dominance and calculate the right-handed Majorana neutrino matrix $M_{RR}$ using the inverse type I seesaw formula 
\begin{equation}
M_{RR}=m_{LR}^Tm_{LL}^{-1}m_{LR}
\label{inverseI}
\end{equation}
To calculate the $M_{RR}$ for each case, we need to have the Dirac neutrino mass matrix $(m_{LR})$. We take the Dirac neutrino mass matrix $m_{LR}$ to be diagonal of the form (\ref{mLR1}). Now, using $m_{LR}$ in the inverse type I seesaw formula above, we calculate $M_{RR}$. We then diagonalize $M_{RR}$ as shown in $(\ref{mrrdiag})$ for the calculation of baryon asymmetry. The diagonalizing matrix of $M_{RR}$ effectively makes $m_{LR}$ non-diagonal as can be seen from $(\ref{mlrdiag})$. Thus, we are studying a model with diagonal $M_{RR}$ and non-diagonal $m_{LR}$. This kind of specific Yukawa structure can be naturally explained by using additional flavor symmetries, either global or gauged. A possible abelian gauge symmetry that can constrain the Dirac neutrino mass matrix to the diagonal form was outlined recently in one of our works (first reference in \cite{mkd-db-rm}). Similar approach can be followed to constrain the right handed neutrino mass matrix to the diagonal form while keeping the Dirac neutrino mass matrix in its general non-diagonal form.

After calculating $M_{RR}$, we calculate the predictions for baryon asymmetry by varying the Majorana CP phases for both normal and inverted hierarchies as well as two choices of Dirac CP phases. Since the calculations for baryon asymmetry are different for different temperature regimes (due to flavor effects), we discuss all the three regimes separately.

\subsection{One Flavor Regime $T > 10^{12} \; \text{GeV}$}
If thermal leptogenesis occur at very high temperature $T > 10^{12}$ GeV, the mass of the lightest right handed neutrino responsible for producing the lepton asymmetry is also as heavy as the temperature scale. Now, from the inverse type I seesaw formula (\ref{inverseI}), the mass of the lightest right handed neutrino depends on the Dirac neutrino mass matrix $m_{LR}$ for a given light neutrino mass matrix $m_{LL}$. The light neutrino mass matrix $m_{LL}$ can be constructed by taking the best fit values of oscillation parameters in neutrino mixing matrix (\ref{matrix1}) and choosing the lightest mass eigenstate value. For normal hierarchy, the diagonal mass matrix of the light neutrinos can be written  as $m_{\text{diag}} 
= \text{diag}(m_1, \sqrt{m^2_1+\Delta m_{21}^2}, \sqrt{m_1^2+\Delta m_{31}^2})$ whereas for inverted hierarchy 
 it can be written as $m_{\text{diag}} = \text{diag}(\sqrt{m_3^2+\Delta m_{23}^2-\Delta m_{21}^2}, 
\sqrt{m_3^2+\Delta m_{23}^2}, m_3)$. The smallest light neutrino mass is chosen in such a way to obey the cosmological upper bound on the sum of absolute 
neutrino masses $\sum_i |m_{i}| < 0.23$ eV \cite{Planck13} reported by the Planck collaboration recently. We choose two possible values of the lightest mass eigenstate $m_1, 
m_3$ for normal and inverted hierarchies respectively. First we choose $m_{\text{lightest}}$ as large 
as possible such that the sum of the absolute neutrino masses fall just below the cosmological upper bound. 
For normal and inverted hierarchies, this turns out to be $0.07$ eV and $0.065$ eV respectively. Then 
we allow moderate hierarchy to exist between the mass eigenvalues and choose the lightest mass eigenvalue 
to be $0.001$ eV to study the possible changes in our analysis and results.

To keep the lightest right handed neutrino mass in this one flavor regime, we need to choose the Dirac neutrino mass matrix with appropriate diagonal entries. We take the same value of lightest active neutrino mass as considered before, and vary $(m,n)$ of $m_{LR}$ given in equation (\ref{mLR1}) to calculate the right handed neutrino masses. We take $m_f = 82.43$ GeV and find that the choice $(m,n)=(1,1)$ keeps the lightest right handed neutrino in the regime $M_1 > 10^{12}$ GeV. We first show our type I seesaw results for baryon asymmetry $Y_B$ with respect to the difference in the Majorana CP phases $\alpha-\beta$ for inverted and normal hierarchies in figure \ref{fig1} and \ref{fig2} respectively. We include both choices of Dirac CP phases $5\pi/3, \pi$ found in global fit data as well as both choices of lightest neutrino mass eigenstate as mentioned above. We then introduce the type II seesaw term as a small perturbation and show the results for baryon asymmetry for the same choices of Dirac CP phase and lightest mass eigenstate (as above) in figure \ref{fig3} and \ref{fig4}. It is to be noted that, here we have taken $\gamma = 1$ in the neutrino mass formula (\ref{type2b}) which corresponds, in our notation, to the smallest contribution of type II seesaw term.

\subsection{Two Flavor Regime $10^9 \; \text{GeV} < T < 10^{12} \; \text{GeV}$}
We again vary $(m,n)$ for a given $m_{LL}$ and find that for $(m,n) = (3,1)$, the lightest right handed neutrino mass is in the range $10^9 \; \text{GeV} M_1 < 10^{12} \; \text{GeV}$ which corresponds to the two flavor regime of leptogenesis as discussed in the previous section. The results for baryon asymmetry as a function of Majorana CP phase difference $\alpha-\beta$ are shown in figure \ref{fig5} and \ref{fig6} by taking into account of type I seesaw. It can be seen from figure \ref{fig5} and \ref{fig6} that for quasi-degenerate type of neutrino mass spectra with lightest mass eigenstate $0.065 \; \text{eV} (\text{IH}), 0.07\; \text{eV} (\text{NH})$ eV, the calculated baryon asymmetry falls short of the observed data. For a hierarchical light neutrino mass spectra with lightest mass eigenstate $0.001$ eV, the observed baryon asymmetry can be reproduced in the model for both normal and inverted hierarchies. We then introduce type II seesaw as a small perturbation and find that the observed baryon asymmetry can be reproduced for both hierarchical and quasi-degenerate light neutrino mass spectra. The results can be seen in figure \ref{fig7} and \ref{fig8}.

\subsection{Three Flavor Regime $T < 10^{9} \; \text{GeV}$}
To keep the lightest right handed neutrino mass below $10^9$ GeV, we take $(m,n)$ as $(5,3)$ in the Dirac neutrino mass matrix (\ref{mLR1}). The calculated baryon asymmetry is shown as a function of Majorana CP phase difference $\alpha-\beta$ in figure \ref{fig9} and \ref{fig10} for type I seesaw. It can be seen that the parameter space giving rise to observed baryon asymmetry decreases significantly as we go from one flavor to three flavor regime. We do not find any parameter space giving rise to observed baryon asymmetry for hierarchical light neutrino masses of inverted type with lightest mass eigenvalue $0.001$ eV. As the lightest neutrino mass eigenvalue is increased to $0.065$ eV, a small parameter space can be found giving rise to correct baryon asymmetry as seen from figure \ref{fig9}. For normal hierarchy on the other hand, we find parameter space (as seen in figure \ref{fig10}) for both choices of lightest neutrino mass $0.001, 0.07$ eV as well as Dirac CP phases $5\pi/3, \pi$. The introduction of type II seesaw term as a perturbation does not alter the parameter space giving rise to correct baryon asymmetry as seen from figure \ref{fig11} and \ref{fig12}.

\section{Results and Conclusion}
\label{conclude}
We have studied the possibility of producing the observed baryon asymmetry in the Universe within the framework of generic left-right symmetric models by considering both hierarchical and quasi-degenerate type light neutrino mass spectra and two different Dirac CP phases ($180^o, 300^o$) found in best-fit global neutrino oscillation data in the literature. In generic LRSM, neutrino mass can get contributions from both type I as well as type II seesaw terms. We first consider only type I seesaw and then consider a case in which type I seesaw dominates and type II seesaw exists as a small perturbation. We use the generic parametrization of neutrino mixing matrix and find the numerical value of these parameters using the global fit neutrino oscillation data. The Majorana CP phases which are currently unconstrained from experimental data, are considered  as free parameters. For the type I seesaw case, we then compute the predictions for baryon to photon ratio as a function of the Majorana CP phases. We then keep the type I seesaw term as the dominant one and introduce the type II seesaw term as a perturbation, the strength of which can be chosen by varying value of the dimensionless parameter $\gamma$. For simplicity, we choose$\gamma = 1$ which corresponds to the minimum contribution of type II seesaw. We then calculate the baryon asymmetry by taking both type I and type II seesaw contribution for both choices of lightest neutrino mass eigenvalue and Dirac CP phase. Our main conclusions can be summarized as follows:
\begin{itemize}
\item For single flavor scenario that is $T > 10^{12}$ GeV, observed baryon asymmetry can be successfully reproduced for both inverted and normal hierarchies and all the choices of lightest neutrino mass eigenvalue and Dirac CP phase. Introducing type II seesaw as a small perturbation do not alter the type I seesaw results substantially.
\item For two flavor scenario $10^9 \; \text{GeV} < T < 10^{12} \; \text{GeV}$, observed baryon asymmetry can not be produced within type I seesaw framework if the light neutrino mass hierarchy is very mild or quasi-degenerate type with the lightest mass eigenstate being $0.065 \; \text{eV} (\text{IH}), 0.07 \; \text{eV} (\text{NH})$. For a hierarchical light neutrino mass spectra with the lightest neutrino mass $0.001$ eV, both normal and inverted hierarchies as well as both the choices of Dirac CP phase can give rise to the correct baryon to photon ratio.
\item Introducing type II seesaw term as a small perturbation can however, reproduce successful baryon asymmetry in the two flavor regime for both choices of lightest neutrino mass eigenstate as well as Dirac CP phases unlike with type I seesaw alone as mentioned above.
\item The parameter space allowing successful leptogenesis in agreement with observation decreases as we go into the three flavor regime $T < 10^9 \; \text{GeV}$. For inverted hierarchy with lightest mass eigenstate $0.001$ eV, we do not get any parameter space giving rise to the observed baryon asymmetry both with type I as well as a combination of type I and type II seesaw. Thus, three flavor regime with inverted hierarchy prefers a very mild hierarchy between light neutrino mass eigenvalues.
\item For normal hierarchy in three flavor regime, both choices of lightest mass eigenvalue and Dirac CP phases can give rise to correct baryon asymmetry with more parameter space for larger value of lightest neutrino mass. Also, the parameter space for type I seesaw does not get altered significantly after the introduction of type II seesaw term for both normal and inverted hierarchies.
\item For all the flavor regimes, we not only put indirect limits on CP phases and lightest neutrino mass eigenvalue but also constrain the Dirac neutrino mass matrix.

\end{itemize}

In view of above, the neutrino mass models considered in our study can survive in nature and produce the observed baryon asymmetry within the framework of type I and type II seesaw mechanism with type I seesaw as the dominating one. For both type I and type II seesaw cases, we can discriminate between possible values of Dirac CP phases, Majorana CP phases as well the patterns of neutrino masses: mild and sizable hierarchies. We have observed that depending on the neutrino hierarchy as well as lightest mass eigenvalue, certain combinations of Dirac and Majorana CP phases are disfavored from successful leptogenesis point of view. Therefore, our results can put certain indirect limits on the values of CP phases which are currently unconstrained from neutrino oscillation experiments. Future determination of CP phases in neutrino experiments should be able to verify or falsify some of the scenarios we have discussed in this work.

\end{document}